\documentclass[a4paper,12pt]{article}
\usepackage{epsfig}% SRC Specials

\newcommand{\be}{\begin{equation}}
\newcommand{\ee}{\end{equation}}
\newcommand{\ba}{\begin{eqnarray}}
\newcommand{\ea}{\end{eqnarray}}
\newcommand{\ban}{\begin{eqnarray*}}
\newcommand{\ean}{\end{eqnarray*}}
\begin{document}

\title{\Large\bf Preferred Frame versus Multisimultaneity:\\
\large\bf meaning and relevance of a forthcoming experiment}

\author{{\bf Antoine Suarez}\thanks{suarez@leman.ch}\\ Center for Quantum
Philosophy\\ The Institute for Interdisciplinary Studies\\ P.O.
Box 304, CH-8044 Zurich, Switzerland}

%\date{}

\maketitle

\vspace{2cm}
\begin{abstract}

It is argued that: 1) Quantum Mechanics implies the preferred
frame also because of the collapse delayed at detection, 2)
forthcoming experiments with moving beam-splitters will allow us
to decide between Preferred Frame and Multisimultaneity, and 3) if
Preferred Frame prevails, superluminal communication is in
principle possible.\\


\end{abstract}

\pagebreak

\section{Introduction: the conflict between Quantum Mechanics
and relativity of simultaneity}

Quantum Mechanics predicts the statistical distribution of
alternative detection events in experiments according to the
superposition principle: in case of a superposition quantum state
the probabilities of the possible outcomes of an experiment have
to be calculated combining the single quantum amplitudes
\cite{feym65}.\\

It is obviously impossible to determine the statistical
distribution of counts an experiment yields other than by counting
single detection events. This is why, even if the quantum formalism
does not bother on single events but on events distributions,
Quantum Mechanics cannot avoid to declare some relationship between
the single detection event and the predicted statistical
distribution. The theory does this through the so-called
``reduction postulate''. A careful analysis shows that this
postulate leads to some assumptions about the mechanism of
measurement \cite{ash89, tm94, whee86, pen89, azsg99}:

\begin{enumerate}
  \item ``Instantaneity of the state-reduction'': a measurement taking
place somewhere affects ``instantaneously'' the whole system
everywhere, and this jumps into the measured eigenstate.

  \item ``Wavefunction collapse at detection'': the outcome of a possible
single measurement is not determined till a detection occurs: as
far as this is not the case the system has to be considered as a
quantum superposition of the possible outcomes, it is at detection
when the ``collapse of the wavefunction'' takes place.

  \item ``External observer'': the detectors themselves can be considered
together with the measured system as a single quantum
``system-apparatus state'' till the whole becomes collapsed by an
external observer.

\end{enumerate}

The relationship between Quantum Mechanics and Relativity has been
object of vast analysis since John Bell showed that: a) if one
only admits relativistic local causality (causal links with $v\leq
c$), the correlations occurring in two-particle experiments should
fulfill clear locality conditions (``Bell's inequalities''), and
b) for these experiments the quantum mechanical superposition rule
bears predictions violating such locality criteria (``Bell's
theorem'') \cite{jb64}. Bell experiments conducted in the past two
decades, in spite of their loopholes, suggest a violation of local
causality: statistical correlations are found in space-like
separated detections; violation of Bell's inequalities ensure that
these correlations are not pre-determined by local events
\cite{aapggr82, jrpt90, ptjr94, tbzg98.1, tbzg98.2, gw98, db98}.
Nature seems to behave non-locally, and Quantum Mechanics predicts
well the observed distributions.\\

Now the question arises: can we use the instantaneous influences
involved in Bell's experiments to built an arbitrarily fast
telephone line? From the point of view of Quantum Mechanics the
answer is a matter of formalism. According to the standard
formalism we cannot use ``Bell influences'' for faster-than-light
communication \cite{eb78, grw80}. Nevertheless it seems that there
is no clear reason why Quantum Mechanics, ``a specifically
nonrelativistic theory'', should prevent superluminal
communication \cite{ghz93}. Effectively nonstandard formalisms of
Quantum Mechanics have been developed which share the predictions
of the standard one for all experiments already done, but lead to
superluminal communication in more sophisticated experiments not
yet performed \cite{eb89, ng90, pcjr99}.\\

The prevention of faster-than-light communication by the standard
formalism led to the celebrated expression of ``peaceful
coexistence'' to characterize the relationship between standard
Quantum Mechanics and Special Relativity \cite{ash89}, and is
often invoked as demonstration that nonlocality does not conflict
with Einstein's relativistic causality. This way of arguing
overlooks somewhat that the principle of Special Relativity
effectively implies that {\em nothing} in nature goes faster than
light, and therefore it is really at odds with the nonlocal
behavior of nature revealed by Bell's experiments. By contrast,
the relativity of simultaneity does not exclude \emph{any}
superluminal influence but only those implying backwards
causation, as for instance influences leading to superluminal
communication between human observers. Hence if one considers that
what actually follows from Michelson-Morley's observations is the
relativity of simultaneity, and therefore this principle (and not
the postulate of Special Relativity) is the essential of
relativity, then it can be appropriately said that there is no
conflict between relativity and the nonlocal ``Bell influences''
so long as these do not lead to superluminal telephone lines or,
more in general, backwards causation.\\

However the way how standard Quantum Mechanics manages nonlocality
bear problems because of the assumed timing independence of the
nonlocal correlations. Hardy showed that if one considers Bell
experiments with moving observers, the quantum mechanical
predictions deriving from the superposition principle conflict
with the relativity of simultaneity and imply the existence of a
preferred frame \cite{lh92, ip98}. In this respect it is
interesting to see that Quantum Mechanics has already been
consistently developed as absolute space-time scheme in the form
of Bohmian Mechanics which keeps instantaneous influences and
disposes of the ``collapse'' \cite{tm94, dbbh, jb87}, or
Eberhard's theory which assumes that the Bell's correlations
originate from signaling at a finite superluminal speed
\cite{eb89}, or Rembieli\'{n}ski's model which remarkably cast
preferred-frame Quantum Mechanics into a Lorentz-covariant scheme
\cite{pcjr99}. As regards other realistic models which incorporate
the ``collapse'' like the GRW theories \cite{grw86, ghirardi90},
since they share the quantum mechanical predictions, Hardy's
theorem implies that such descriptions also lead in principle to a
preferred-frame.\\

Concerning the ``reduction postulate'' Aharonov and Albert did
argue that, if one assumes the impossibility of a preferred frame,
the assumption of  ``instantaneity'' requires to give up the very
concept of state \cite{yada81}. As regards the collapse of the
``system-apparatus state''  by the ``external observer'', Peres
did claim this notion ``to have no meaning whatsoever in a
relativistic context'' \cite{pe95}.\\

In summary, the ``quantum superposition principle'', the
``instantaneity of the state-reduction'' and the
``system-apparatus state'' have been considered to be at odds with
the relativity of simultaneity. By contrast, to our knowledge, the
assumption that the outcome is not determined till detection has
never been suspected of conflicting with it. In this article we
argue that the assumption of collapse at detection excludes in
principle the relativity of simultaneity, so that Quantum
Mechanics has to be considered a preferred-frame theory because of
all its specific features. This conclusion let appear experiments
with moving beam-splitters as particularly relevant in order to
decide whether nature itself really uses a preferred-frame in
working out the phenomena. Would this be the case, then nothing in
principle speaks against the possibility of superluminal
signaling, and the lower bound on the ``speed of quantum
information'' recently set by experiment \cite{zbtg99} would
acquire practical interest.

\section{The unification of nonlocality and relativity
into Multisimultaneity}

In light of the predicting success of Quantum Mechanics and Hardy's
theorem, it is tempting to think that: ``although nonlocality does
not require a special frame of reference, it is most naturally
incorporated into a theory in which there is a special frame of
reference'' \cite{lh92}.\\

Recent work has proposed a different line of thinking
\cite{asvs97.1, as97.2, as98.1}: accepting superluminal
nonlocality and relativity of simultaneity as experimental facts,
one has to modify to some extent the rule establishing when the
probabilities are calculated by summing of amplitudes. The result
is a many frames description called Multisimultaneity or
Relativistic Nonlocality, which makes predictions contradicting
Quantum Mechanics in the context of experiments not yet performed.
The new description shows that if one assumes Nature itself to use
many frames to cause the phenomena (i.e. {\em real} relativity
instead of the conventional observer's one), then superluminal
nonlocality can become most naturally incorporated into a
many-frames theory.\\

Multisimultaneity gives up the key role Quantum Mechanics
attributes to detection and ``collapse'', as Bohmian Mechanics
also does. What causes a detector to fire, the ``observable
particle's part'' (or simply the ``particle''), always travels a
definite path, but undetectable information (similarly to the
Bohmian ``empty'' wave) does travel the alternative paths. One has
two kinds of events: The basic event is the ``choice'' of the path
the ``particle'' takes at the beam-splitters, which for this
reason are called the ``choice devices''. The second event is
detection, which simply reveals the channel by which the
``particle'' left the preceding ``choice device'', but produces
``irreversibility'' in the sense that after detection it is no
longer possible to arrange that the ``particle'' and the
unobservable information (traveling the other channel) meet
again.\\

In the context of two-particle experiments, Multisimultaneity is
implemented as follows: at the instant $(T_{i1})_{i1}$ particle
$i$ meets the beam-splitter BS$_{i1}$ it is considered whether in
the referential frame of this device particle $j$ did already meet
BS$_{j1}$ (i.e. whether ($T_{j1}\leq T_{ik})_{ik}$), and, in case
of several alternative paths, whether it is impossible to
distinguish by which path pair the particles did enter BS$_{i1}$
and BS$_{j1}$ on the basis of any possible experiment allowing us
to monitor the output ports of BS$_{i1}$ and BS$_{j1}$. If these
two conditions are met particle $i$ produces its outcomes taking
account of the phase parameters particle $j$ meets at the other
side of the setup, and if not particle $i$ produces its outcomes
without taking account of the phase parameters particle $j$ meets.
A particular two-particle experiment is that with {\em
before-before} timing \cite{asvs97.1}, in which each particle
chooses only according to local parameters, and for which
Multisimultaneity predicts the absence of nonlocal correlations.
Thus, regarding the predictions for new experiments with
beam-splitters in motion Multisimultaneity deviates from Quantum
Mechanics.\\

\section{Quantum collapse at detection excludes relativity of simultaneity}

The fact that the detection outcome becomes determined when the
particle meets the monitored ``choice-device'' means, in
particular, that to test Multisimultaneity vs Quantum Mechanics in
Bell experiments with two referential frames, one has to set the
beam-splitters in motion to generate {\em 2-before} impacts.
Nevertheless, at the beginning of the work to prepare these
experimental tests the question arose, whether a version of
Multisimultaneity keeping the ``collapse at detection'' is
possible, and one should consider that the detectors are the
actual ``choice-devices'' at which the outcomes become determined.
The question was relevant because depending on the answer one has
to set the detectors in motion, instead of the beam-splitters. As
said above, nothing in principle seemed to speak against the
incorporation of ``collapse at detection'' into a many frames
theory. The fact that Bohmian Mechanics, a proper non-relativistic
theory, disposes of ``collapse at detection'' let rather appear as
plausible such an incorporation.\\

However it is possible to show that as far as one keeps to the
basic `one photon-one count' principle, the assumption of collapse
at detection excludes in principle the relativity of simultaneity,
and, therefore, implies the preferred frame \cite{as98.1}. In the
following we give a proof that this already holds for single
particle experiments, and therefore the incompatibility of
collapse and relativity of simultaneity is deep rooted in Quantum
Mechanics.\\

%%%%%%%%%%%%%%%%%%%%%%%%%%%%%%
\begin{figure}[t]
\centering\epsfig{figure=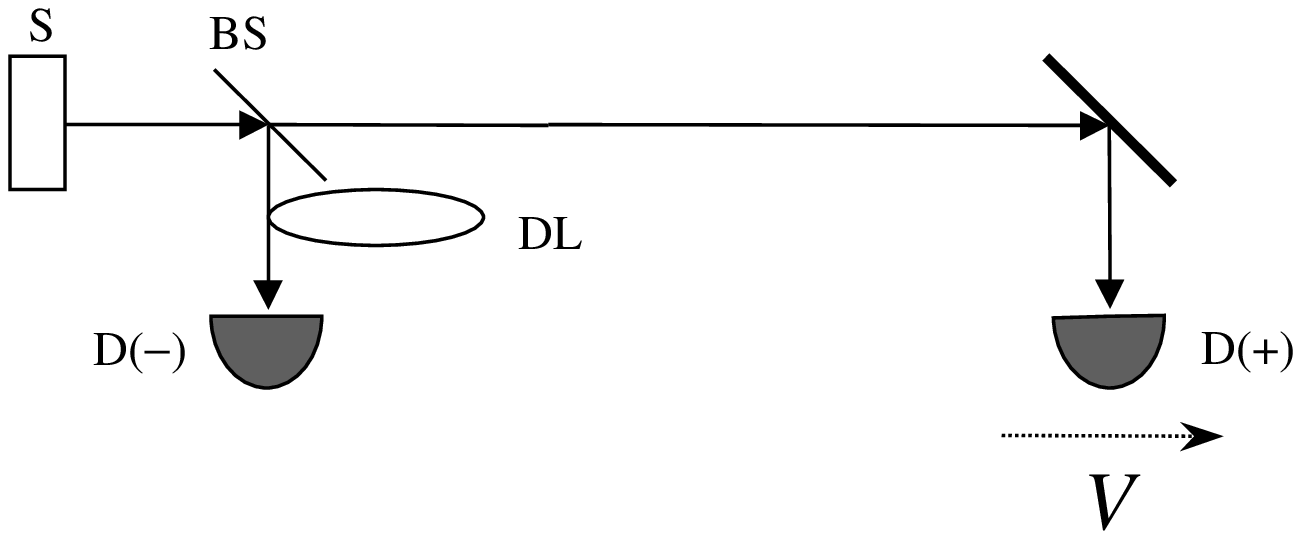,width=120mm}
{\small{\it{\caption{Single-particle gedankenexperiment assuming
that the arrival of the wave at each D$(\sigma),\sigma
\in\{+,-\}$, occurs, in the referential frame of D$(\sigma)$,
before the arrival of the wave to D$(-\sigma)$.}}}} \label{ccfig1}
\end{figure}
%%%%%%%%%%%%%%%%%%%%%%%%%%%%%%

Consider the experiment sketched in Fig. 1 in which single photons
emitted from source S impact into a 50-50 beam-splitter BS and get
detected thereafter either in detector D$(+)$, or D$(-)$. By means
of delay line DL the optical paths are adjusted so that the
arrival of the wave at D$(-)$ occurs in the laboratory frame a
little bit before the arrival D$(+)$. Suppose the two detectors
far away from each other so that a light signal sent at D$(-)$ at
the instant of the arrival of the wave at D$(-)$, cannot reach
D$(+)$ before the arrival of the wave at D$(+)$.\\

If detectors are the choice-devices, which detector fires is not
determined before the wave reaches the detectors. But if at this
instant only one of the detectors can fire, there must be some
kind of superluminal influence (``Bell connection'') between the
two detectors. According to the basic principle of
Multisimultaneity, at the arrival of the wave at each D$(\sigma)$,
$\sigma \in\{+,-\}$, the choice between ``fire'' and ``not-fire''
of D$(\sigma)$ depends on whether, in this device's frame, a
choice in D$(-\sigma)$ did already take place or not. Suppose
first both detectors at rest in the laboratory frame: since at the
time of arrival at D$(-)$ the wave did not yet reach D$(+)$, then
the choice in D$(-)$ takes place at random; by contrast the choice
at D$(+)$ takes account of the choice at D$(-)$: if D$(-)$ did
fire, then D$(+)$ does not fire, and reversely.\\

Suppose now the detector D$(+)$ in motion so that the arrival of
the wave at each D$(\sigma)$ occurs, in the referential frame of
D$(\sigma)$, before the arrival at D$(-\sigma)$. This can be
ensured by fulfilling the same condition given in \cite{asvs97.1}
to produce two {\em before} impacts at the beam-splitters. Then
the fact that each D$(\sigma)$ fires, cannot depend on whether
D$(-\sigma)$ fires or not, and therefore it should happen that,
even if there is one and only one particle traveling the setup,
$25\%$ of the times D$(+)$ and D$(-)$ fire jointly, and $25\%$ of
the times neither D$(+)$ nor D$(-)$ fires. This means a violation
of the basic assumption that one single photon cannot cause two
detectors to fire. Therefore, as far as one keeps to the `one
photon-one count' principle, the quantum collapse excludes the
relativity of simultaneity.

\section {An imminent experiment will allow us to decide between
Preferred Frame and Multisimultaneity}

The proof of the preceding section means that also the collapse
delayed at detection makes of Quantum Mechanics a preferred-frame
theory. Therefore, as far as one wishes avoid backwards causation
\cite{co97} and keeps the principles of causality and `one
photon-one count', Quantum Mechanics has to be considered a
preferred-frame theory because of all its basic ideas. In this
sense the real status of the GRW theory \cite{jb87, grw86} seems
to be that of a preferred-frame quantum theory with ``quantum
collapse''.\\

Regarding superluminal Bell influences (nonlocality), as already
noticed in Section 2, they can be incorporated as well into
preferred-frame Quantum Mechanics, as many-frames
Multisimultaneity.\\

In light of this analysis the upcoming experiment with moving
beam-splitters \cite{as00.1} acquires special relevance: it will
test two types of nonlocal descriptions against each other and
allow us to decide between preferred-frame and many-frames
theories, similarly as Bell experiments allow us to decide between
nonlocal and local ones.

\section {Assumed the preferred frame, nothing in principle speaks against
superluminal communication}

Vindication of Quantum Mechanics by the experiment with moving
beam-splitters referred to would mean that one has to give up the
relativity of simultaneity and accept the preferred-frame. A
preferred-frame description could be done as well without
superluminal signaling in the form of the Bohm's theory (without
collapse) \cite{dbbh} or a GRW theory (with collapse) \cite{grw86,
jb87}, as with superluminal signaling in the form for instance of
Eberhard's theory \cite{eb89}, Weinberg's non-linear Quantum
Mechanics \cite{ng90}, or even Lorentz-covariant schemes
\cite{pcjr99}.\\

Nevertheless we would like to stress that if one accepts the
preferred frame the impossibility of superluminal communication
can neither originate from special relativity nor from causal
requirements related to relativity of simultaneity (to avoid
backwards-causation), so that it is not clear in name of which
``Relativity'' such an impossibility could still be maintained
\cite{gs99}: in any case, within the preferred frame
faster-than-light propagation of information is consistent with
causality \cite{pcjr99}.\\

And as regards Quantum Mechanics itself, let us say once again,
that the whole issue is a pure matter of formalism and not of
principle: if ``incompatibility with relativity'' cannot longer be
invoked, ``supraluminal communication'' cease to be an argument
against ``non-linear Quantum Mechanics'' \cite{ng90}, and in the
context of an absolute space-time scheme it becomes quite
reasonable to explain the correlations appearing in Bell
experiments by means of a finite superluminal speed \cite{eb89}.\\

What is more, for reasons of experimental consistency, to exclude
a conflict with Michelson-Morley-like observations,
preferred-frame schemes have to assume faster-than-light
propagation of energy over open paths \cite{pcjr99}. Consequently
it would be only natural that the formalism of preferred-frame
Quantum Mechanics exploits such a possibility of superluminal
communication.\\

All this means the following: if the imminent experiment using
moving beam-splitters \cite{as00.1} uphold the preferred-frame
description, then one should seriously take into consideration the
possibility of realizing superluminal communication by means of
more sophisticated experimental arrangements; maybe corresponding
gedankenexperiments already proposed within different nonstandard
formalisms \cite{eb89, ng90, gs99} can be a source of valuable
inspiration. In such a context also the Gisin-Zbinden lower bound
on the ``speed of quantum information'' set by experiment
\cite{zbtg99} would acquire the new strong meaning that telephone
lines faster than ``10 million times the velocity of light'' are
\emph{in principle} possible.

\section {Conclusion}

Quantum Mechanics appears to be a preferred-frame theory not only
because of the predictions deriving from the superposition
principle (Hardy's theorem) \cite{lh92}, but also because of the
collapse delayed at detection. Vindication of Quantum Mechanics by
experiments with moving beam-splitters would mean that one has to
accept the Preferred Frame. Within this context nothing speaks
against considering that communication for practical purposes may
be possible at velocities of 10 million times the velocity of
light. To think that such velocities of communication are
``unrealistic'', means in fact to share our conviction that the
Preferred Frame is not the correct view, and Quantum Mechanics
will fail in the coming experiments with beam-splitters in motion
\cite{as00.1}. In conclusion, a unique experimental result is
expected within this year: either Multisimultaneity holds and
Quantum Mechanical fails, or the Preferred Frame prevails and
superluminal communication is in principle possible. \\

\section*{Acknowledgements}

I am indebted to Nicolas Gisin, Jakub Rembieli\'{n}ski, Valerio
Scarani, Gao Shan and Hugo Zbinden for very useful suggestions and
communications, and thank Ian Percival, Sandu Popescu, and John
Rarity for discussions during the Workshop on Relativistic
Nonlocality (GAP-Fondation Odier, Geneva, November 1998). I also
acknowledge support by the L\'eman Foundation and the Odier
Foundation for psycho-physics.

\end{document}